\documentclass[aps,twocolumn,pra,tightenlines,floatfix,showpacs]{revtex4}
\usepackage[dvips]{graphicx}
\usepackage[english]{babel}
\usepackage{amsmath}
\usepackage{amssymb}
\usepackage{times}

\newcommand{\bs}{\begin{split}}
\newcommand{\es}{\end{split}}
\newcommand{\be}{\begin{equation}}
\newcommand{\ee}{\end{equation}}
\newcommand{\ba}{\begin{eqnarray}}
\newcommand{\ea}{\end{eqnarray}}
\newcommand{\ek}{\epsilon_{\mathbf{k}}}
\newcommand{\Ek}{E_{\mathbf{k}}}
\newcommand{\uk}{u_{\mathbf{k}}}
\newcommand{\vk}{v_{\mathbf{k}}}
\newcommand{\sumk}{\sum_{\mathbf{k}}}

\newcommand{\sumq}{\sum_{\mathbf{q}}}

\newcommand{\Omegaq}{\Omega_{\mathbf{q}}}

\begin{document}

\title{Intermediate temperature superfluidity in an atomic Fermi gas
  with population  imbalance}

\author{Chih-Chun Chien, Qijin Chen, Yan He, and K. Levin}

\affiliation{James Franck Institute and Department of Physics,
 University of Chicago, Chicago, Illinois 60637}

\date{\today}

\begin{abstract}
  We derive the underlying finite temperature theory which describes
  Fermi gas superfluidity with population imbalance in a homogeneous
  system.  We compute the pair formation temperature and superfluid
  transition temperature $T_c$ and superfluid density in a manner
  consistent with the standard ground state equations, and thereby
  present a complete phase diagram.  Finite temperature stabilizes
  superfluidity, as manifested by two solutions for $T_c$, or by low $T$
  instabilities.  At unitarity the polarized state is an ``intermediate
  temperature superfluid".
\end{abstract}

\pacs{03.75.Hh, 03.75.Ss, 74.20.-z \hfill \textsf{Journal Ref: Phys. Rev. Lett. \textbf{97}, 090402 (2006)}}

\maketitle

Excitement in the field of ultracold Fermi gases has to do with their
remarkable tunability.  As a magnetic field is varied from weak to
strong this system undergoes a transition from a Bose Einstein condensate
(BEC) to a BCS-based superfluid.  Recently \cite{ZSSK06,PLKLH06} another
tunability has emerged; one can vary the population imbalance between
the two ``spin" species.  This capability has led to speculations about
new phases of superfluidity, quantum critical points, and has
repercussions as well in other sub-fields of physics \cite{LW03,FGLW05}.
A nice body of theoretical work on this subject
\cite{PWY05,SM06,HS06,PS05a} has focused on zero temperature ($T$)
studies of the simplest mean field wavefunction \cite{Leggett}, with
population imbalance.  Additional important work presents \cite{SR06} a
$T=0$ two-channel, mean-field approach for very narrow Feshbach
resonances, as well as a study of finite $T$ effects \cite{YD05}, albeit
without a determination of superfluid order.

Superfluidity is, generally, a finite $T$ phenomenon and it is the
purpose of the present paper to explore finite temperature effects
\cite{ourreview,Kosztin1,Chen2} based on the BCS-Leggett ground state
 with population imbalance.  We determine the behavior of the pair
formation temperature $T^*$, the superfluid transition temperature,
$T_c$, and superfluid density $n_s(T)$, and, thereby arrive at a phase
diagram which addresses general $T$.  Importantly, we find that in the
fermionic regime, superfluidity exists at finite $T$ (although not at
$T=0$), leading to the new concept of an ``intermediate temperature
superfluid".  Because temperature acts in this rather unexpected
fashion, 
we reduce the complexity and confine our attention to the homogeneous
system.

The approach which we outline below, importantly, includes what we call
``pseudogap effects". 
For $T \neq 0$, the excitation gap
$\Delta$ is different from the order parameter, 
due to the contribution to $\Delta$ from \textit{noncondensed pairs}
\cite{ourreview,PS05}.  In this way, the solution 
for $T_c$ is
necessarily different from that obtained in the literature.  Generally,
$\Delta^2$ contains two additive contributions \cite{Kosztin1,Chen2}
from the condensate ($\Delta_{sc}^2$) and noncondensed pairs
($\Delta^2_{pg}$), and they are proportional to the total, condensed,
and noncondensed pair densities, respectively. This decomposition is 
analogous to the particle number constraint in ideal BEC.  We
emphasize that the central equations derived below are \textit{not}
compatible \cite{PS05} 
with
 the $T \neq 0$ formalism of Ref.~\cite{NSR}.  
In addition, the ``naive mean field theory" with the unphysical
assumption that $ \Delta(T) \equiv \Delta_{sc}(T)$ is not a correct
rendition of $T \neq 0$ effects associated with the BCS-Leggett ground
state.

We define the noncondensed pair propagator, as $t(Q)= U/[1+U\chi(Q)]$,
where, as in Ref.~\cite{Chen2}, the pair susceptibility, given by
$\chi(Q) = \frac{1}{2} \sum_{K} [G_{0\uparrow}(Q-K)G_{\downarrow}(K) +
G_{0\downarrow}(Q-K)G_{\uparrow}(K)]$, can be derived from equations for
the Green's functions , consistent with the BCS-Leggett ground state
equations.  Here $G_\sigma(K)$ and $G_{0,\sigma}(K)=i\omega_n
-\xi_{\mathbf{k},\sigma}$ are the full and bare Green's functions (with
$\sigma = \uparrow, \downarrow$, $\xi_{\mathbf{k},\sigma} =
\epsilon_\mathbf{k} -\mu_\sigma$).  We adopt a one-channel approach
since the $^6$Li resonances studied thus far are broad and consider a
Fermi gas of two spin species with kinetic energy $\epsilon_\mathbf{k} =
\hbar^2 k^2/2m$ and chemical potential $\mu_\uparrow$ and
$\mu_\downarrow$, subject to an attractive contact potential ($U<0$)
between the different spin states.
We take $\hbar=1$, $k_B=1$, and $K\equiv (i\omega_n,
\mathbf{k})$, $Q\equiv (i\Omega_n, \mathbf{q})$, $\sum_K \equiv T\sum_n
\sum_{\bf k}$, etc, where $\omega_n (\Omega_n)$ is the standard odd
(even) Matsubara frequency.

In the superfluid state, the ``gap equation" is 
given by
$U^{-1}+\chi(0)=0$, which is 
equivalent to $\mu_{pair}=0$, the BEC condition of the pairs.  Below
 $T_c$ the self-energy can be well approximated \cite{ourreview} by the
 BCS form, $\Sigma_\sigma(K) = -\Delta^2 G_{0,\bar{\sigma}}(-K)$, where
 $\bar{\sigma} = -\sigma$.  Therefore, $G_{\uparrow,\downarrow}(K) =
 u_\mathbf{k}^2/(i\omega_n \pm h -\Ek) + v_\mathbf{k}^2/(i\omega_n \pm h
 +\Ek) $, where $\Ek = \sqrt{\xi_\mathbf{k}^2 +\Delta^2}$,
 $\xi_\mathbf{k} = \ek - \mu$, $\mu=(\mu_\uparrow + \mu_\downarrow)/2$,
 $h = (\mu_\uparrow - \mu_\downarrow)/2$, and $u_\mathbf{k}^2,
 v_\mathbf{k}^2 = (1\pm \xi_\mathbf{k}/\Ek)/2$. Since the polarization
 $p>0$, we always have $h > 0$.

The ``gap equation" can then be rewritten in terms of the
two-body $s$-wave scattering length $a$, leading to 
\begin{equation}
\label{eq:geq}
\frac{m}{4\pi
  a}=\sum_{\mathbf{k}}\left[\frac{1}{2\epsilon_{\mathbf{k}}}-\frac{1-
    2\bar{f}(E_{\mathbf{k}})}{E_{\mathbf{k}}} \right]. 
\end{equation}
where $ \bar{f}(x) \equiv [f(x+h)+ f(x-h)]/2 $, and 
$ m/4\pi a=1/U+\sum_{\mathbf{k}}(2\epsilon_{\mathbf{k}})^{-1}$. Here
$f(x)$ is the Fermi distribution function.
We define $pn \equiv \delta n =
n_\uparrow -n_\downarrow >0$, where $n= n_\uparrow +n_\downarrow$ is the
total atomic density, and $p=\delta n/n$.
%
Similarly, using $n_{\sigma}=\sum_{K}G_{\sigma}(K)$, one can write 
\begin{subequations}
\label{eq:neq}
\begin{eqnarray}
\label{eq:neqa}
n &=& 2\sum_\mathbf{k} \left[\vk^2 + \frac{\xi_\mathbf{k}}{\Ek}
  \bar{f}(\Ek)\right],\\
pn &=& \sum_\mathbf{k} [f(\Ek-h)-f(\Ek+h)]
\label{eq:neqb}
\end{eqnarray}
\end{subequations}
%
%
Note that, except for the number difference 
[Eq.~(\ref{eq:neqb})], all equations including those below
can be obtained from their unpolarized counterparts by replacing $f(x)$
and its derivative $f'(x)$ with $\bar{f}(x)$ and $\bar{f}'(x)$,
respectively.

While Eqs.  (\ref{eq:geq})-(\ref{eq:neq}) have been written down in the
literature \cite{PWY05,YD05}, the present derivation can be used to go
further and to determine the dispersion relation 
and the number density 
for \textit{noncondensed} 
pairs.
We find
\begin{equation}
  \Delta_{pg}^{2}\equiv - \sum_{Q\neq 0}t(Q) \,,
\label{eq:pgeq}
\end{equation}
which vanishes at $T=0$, where $\Delta^2 = \Delta_{sc}^2$.
In the superfluid phase, $t^{-1}(Q)= \chi (Q)-\chi(0)\approx
Z(\Omega-\Omegaq)$ to first order in $\Omega$, and after
analytical continuation ($i\Omega_n\rightarrow\Omega + i 0^+$). 
Here $\chi(Q) =  \sum_{\mathbf{k}}
   \left[ \frac{1-\bar{f}(E_{\mathbf{k}})
      -\bar{f}(\xi_{q-k})
    } {\Ek+\xi_\mathbf{q-k} - i\Omega_n}\uk^2  -
  \frac{\bar{f}(E_{\mathbf{k}}) -\bar{f}(\xi_{q-k})}
  {\Ek-\xi_{q-k}+i\Omega_n} \vk^2 \right]$.
It follows that
the inverse residue $Z= [ n-2 \sumk \bar{f}(\xi_\mathbf{k})]/2\Delta^2$.
 Thus $\Delta_{pg}^{2}=Z^{-1}\sumq b(\Omegaq)$, where $b(x)$ is the Bose
 distribution function.  To lowest order in $\mathbf{q}$, the pair
 dispersion $\Omegaq = q^2/2M^*$, where the effective pair mass $M^*$
 can be computed from a low $q$ expansion of $\Omegaq$.  
This $q^2$ dispersion is associated \cite{ourreview} with BCS-type
 ground states, which have been the basis for essentially all population
 imbalance work.

Importantly, Eqs.  (\ref{eq:geq})-(\ref{eq:pgeq}) can be used to
determine $T_c$ as the extremal temperature(s) in the normal state at
which noncondensed pairs exhaust the total weight of $\Delta^2$ so that
$\Delta_{pg}^2 = \Delta^2$.  Solving for the ``transition temperature"
in the absence of pseudogap effects leads to the quantity $T_c^{MF}$.
More precisely, $T_c^{MF}$ is defined to be the temperature at which 
 $\Delta(T)$ vanishes within Eqs.~(\ref{eq:geq}) and
(\ref{eq:neq}).  This provides a reasonable estimate for the pairing
onset temperature $T^*$, when a stable superfluid phase exists.  It
should be noted that $T^*$ represents a smooth crossover rather than an
abrupt phase transition, and that Eq.~(\ref{eq:geq}) must be altered
\cite{ChenThermo} above $T_c$ to include finite $\mu_{pair}$.  We will
see that understanding the behavior of $T_c^{MF}$ is a necessary first
step en route to understanding the behavior of $T_c$ itself.

The superfluid density $n_s(T)$ is also required to vanish at the same
value(s) for $T_c$, as deduced above.  Our calculation of $n_s$ closely
follows previous work \cite{Chen2,JS} for the case of the unpolarized
superfluid.  There is an important cancellation between the current
vertex and self-energy contributions involving $\Delta_{pg}^2$ so that,
as expected, $n_s(T)$ varies with the order parameter $\Delta_{sc}^2$.
It is given by
\begin{equation}
\label{eq:n_s}
n_s (T) = \frac{4}{3}\Delta_{sc}^2 \sum_{\bf k}
\frac{\ek}{\Ek^2}
\left[\frac{1-2\bar{f}(E_{\bf{k}})}{2E_{\bf{k}}} + \bar{f}'(E_{\bf{k}})
\right] \,.
\end{equation}
which at $ T =0 $ agrees with Ref.~\cite{PWY05}.

The stability requirements for the superfluid phase have been discussed
in the literature \cite{PWY05}. In general, one requires that the
superfluid density be positive and that the 2x2 ``number susceptibility"
matrix for $ \partial n_{\sigma} / \partial \mu_{\sigma '} $ have only
positive eigenvalues when the gap equation is satisfied.  The $\Delta$
dependence of $n_\sigma$ introduces into the matrix the overall factor
$\left( \frac{\partial^2 \Omega}{\partial \Delta^2
}\right)_{\mu,h}^{-1}$.  Thus, the second stability requirement is
equivalent to the condition that
\begin{equation}\label{eq:d2F}
 \left( \frac{\partial^2 \Omega}{\partial \Delta^2 }\right)_{\mu,h} = 2\sum_{\mathbf{k}}
  \frac{\Delta^2 } {\Ek^2}
  \left[ \frac{1-2\bar{f}(\Ek)} {2\Ek} + \bar{f}^\prime(\Ek) \right] > 0.
\end{equation}
Here $\Omega$ is the thermodynamical potential.  
A third stability requirement, specific to the present calculations, is
that the pair mass $M^* >0$.

\begin{figure}
\centerline{\includegraphics[clip,width=3.2in]{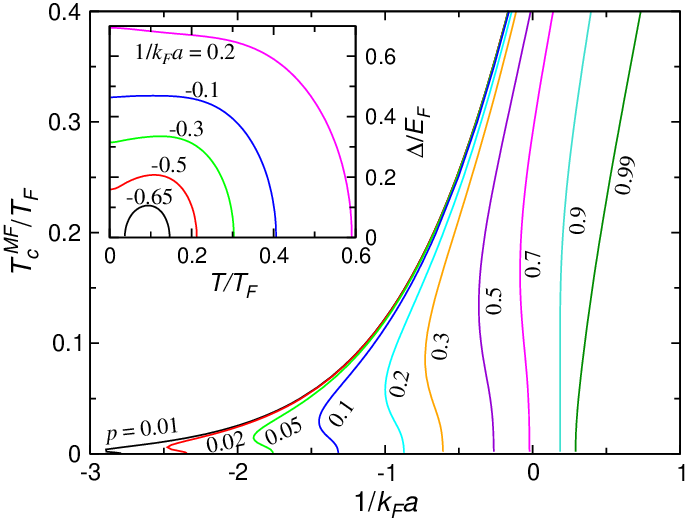}}
\caption{(Color online) Mean-field behavior of $T_c^{MF}$ as a function
  of $1/k_Fa$ for different 
$p$.  Shown in the inset is the pairing gap $\Delta(T)$ at different
  $1/k_Fa$ for $p=0.3$.  Here $E_F\equiv k_B T_F \equiv\hbar^2 k_F^2/2m$
  is the noninteracting Fermi energy 
for $p=0$.
}
\label{fig:TcMF}
\end{figure}

In Fig.~\ref{fig:TcMF}, we present a plot of $T_c^{MF}$ as a function of
 $1/k_Fa$ for a range of $p$.  In the inset
 we plot $\Delta(T)$ at different $1/k_{F}a$ for
$p=0.3$.  For $p<0.9$ and sufficiently low $T_c^{MF}$, the curves for
$T_c^{MF}$ develop an unexpected structure, as one sweeps toward the BCS
regime.  Once $1/k_{F}a$ is less than a critical value, $(1/k_{F}a)_c$
where $T_c^{MF}$ vanishes, there are two $T_c^{MF}$ lines. The lower
branch starts from $(1/k_{F}a)_c$ and increases as $1/k_{F}a$ decreases.
This structure implies that $\Delta$ is nonmonotonic \cite{Sedrakian} in
$T$, as indicated by the bottom curve in the inset of
Fig.~\ref{fig:TcMF}. The two zeroes of $\Delta$ represent the two values
of $T_c^{MF}$.  In contrast to the more conventional behavior (shown in
the top curve for stronger pairing interaction), $\Delta$
\textit{increases} with $T$ at low temperature when $1/k_{F}a$ is
sufficiently small.  This indicates that \textit{temperature enables
pairing}. This was also inferred in Ref.~\cite{YD05}.  In general
superfluids, one would argue that these two effects compete.

\begin{figure}
  \centerline{\includegraphics[clip,width=3.2in] {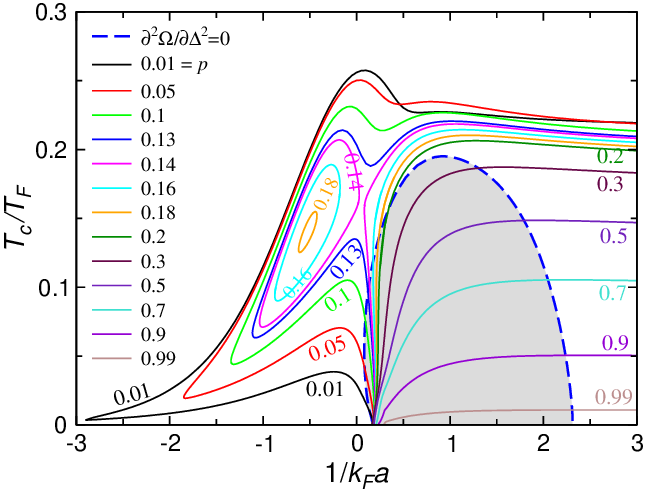}}
  \caption{(Color online) $T_c$ as a function of $1/k_Fa$ for different
 $p$. The $T_c$ curve splits into two disconnected
    curves for $0.14 \lesssim p \lesssim 0.185$.  The $T_c$ solution
    inside the shaded area is unstable.  }
\label{fig:Tc1}
\end{figure}

Insight into this important phenomenon in the fermionic regime
($\mu > 0$), is provided by studying the
momentum distribution $n_\sigma(k)$ at $T=0$ and finite $T$.
 At
$T=0$, pairing is present only for $\ek$ below $\epsilon_{1}=
\textrm{Max}(0,\mu-\sqrt{h^{2}-\Delta^{2})}$ and above $\epsilon_{2}=
\mu+\sqrt{h^{2}-\Delta^{2}}$.  This polarized $T=0$ state requires that
pairs persist to relatively high energies
$\epsilon_{\mathbf{k}}>\epsilon_{2}$, as a result of the Pauli principle
which pushes these states out of the ``normal" regime occupied by the
majority species. This kinetic energy cost competes with the gain from
condensation energy and for sufficiently weak attraction this ``breached
pair" structure \cite{Sarma63} becomes unstable at $T=0$.  By contrast,
at finite $T$ the regime originally occupied exclusively by the majority
species between $\epsilon_1$ and $\epsilon_2$ is no longer completely
filled and pairs can ``spill over" from both lower and higher energy
states into this regime.  This not only helps lower the kinetic energy
but allows the ``normal" regime to participate in pairing and thus
lowers the potential energy.  In this way temperature can enhance
pairing. It should be noted that the majority species between
$\epsilon_1$ and $\epsilon_2$ contains a pairing self-energy and is
different from a free Fermi gas.

Figure \ref{fig:Tc1} represents solutions for $T_c$ of our central
equation set [Eqs.  (\ref{eq:geq})-(\ref{eq:pgeq})] as a function of
$1/k_{F}a$ for the entire range of $p$. 
If the solution for $T_c$ falls into the shaded region, there is
no stable superfluid 
(since
$\partial^2 \Omega/\partial\Delta^2 < 0$, through Eq. (\ref{eq:d2F})).  
For low polarizations $p\lesssim 0.185$, the behavior of $T_c$ is
similar to that of $T_c^{MF}$ when one approaches the BCS regime. There
may be one or two $T_c$'s which, when stable, will be associated with
intermediate temperature superfluidity. When $p > 0.185$, however, no
solution can be found for the regime $1/k_Fa \lesssim 0.18$, because
$M^* <0$ there.  We stress that \textit{the origin of the intermediate
temperature superfluid we find here lies in a very early stage of the
calculations}; it can already be seen as a consequence of the
constraints imposed on the \textit{pairing gap} in the low $T$ regime
when there is a delicate energetic balance between normal and paired
states [see, Eqs.~(\ref{eq:geq}) and (\ref{eq:neq}), and,
Fig.~\ref{fig:TcMF}].

\begin{figure}
\centerline{\includegraphics[clip,width=3.2in]{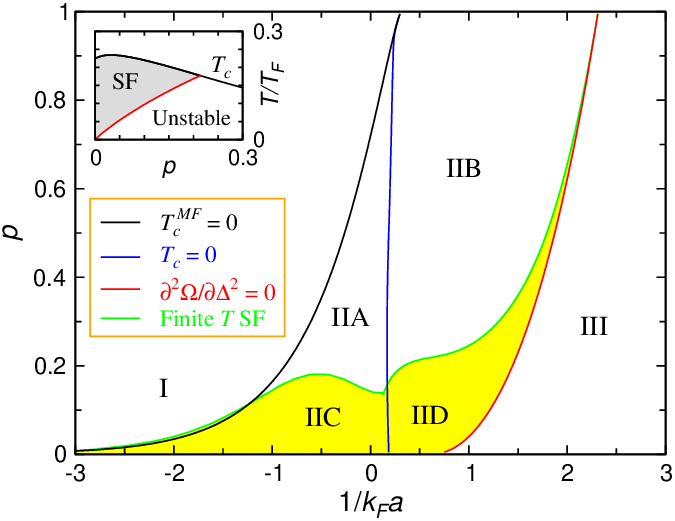}}
\caption{(Color online) Phase diagram on the $p$ -- $1/k_Fa$ plane with
  nearly vertical line 
corresponding to $T_c =0$. Yellow region (shaded) corresponds to
  intermediate temperature superfluidity. Region I the system is normal;
  IIA, $M^* <0$ so no solution for $T_c$ exists; IIB: Solution for $T_c$
  exists but superfluid state is unstable; IIC: Superfluid state exists
  at intermediate $T$ between upper and lower $T_c$'s but not $T=0$;
  IID: Superfluid state exists at finite $T$ but becomes unstable at a
  low temperature $T_{unstable}$ (shown in the inset for $1/k_Fa = 0.5$)
  where $n_s$ is finite; III: Superfluid state exists for all $T\le
  T_c>0$. The chemical potential $\mu$ changes sign within regions IIB
  and IID (close to $1/k_Fa \approx 0.6$ for all $p$), while $n_s(0)$
  vanishes along a nearly vertical line between $(p, 1/k_Fa) = (1, 0.3)$
  to $(0, 0.6)$.  The $p\equiv 0$ boundary is not continuously connected
  to the rest of the phase diagram.  }
\label{fig:phase}
\end{figure}

In Fig.~\ref{fig:phase} we summarize our observations in the form of a
general temperature phase diagram. In region I, the system is normal and
superfluidity is absent.  However, this normal phase need not be a Fermi
gas. Close to the boundary, as shown in the inset to Fig.~\ref{fig:TcMF}
(bottom curve) finite $T$ pairing may occur with or without phase
coherence. In region III, stable superfluidity is present for all $T \le
T_c$.  Finally in region II, within the shaded region (IIC and IID), we
find a stable polarized superfluid phase for intermediate temperatures,
not including $T=0$, which we refer to as intermediate temperature
superfluidity.  In IIA and IIB no stable polarized superfluid is found.
The nearly vertical blue line shown in the figure represents the line
$T_c =0$ which appears around $1/k_{F}a \approx 0.18$ (See
Fig.~\ref{fig:Tc1}) and is roughly independent of polarization.

Finite momentum condensates \cite{Machida05,SR06} may well
occur in any of the regimes in II, particularly IIA and IIB for which
our equations do not yield stable zero momentum condensation.
Future work will explore the nature of the stable phases in these
regimes.
The boundaries of the region denoted II can be compared with other $T=0$
phase diagrams in the literature \cite{SR06,PWY05}.  In contrast to Ref.
\cite{PWY05}, we find that the most stringent criterion for stability at
$T=0$ is 
the positivity of the second order partial derivative of 
$\partial^2 \Omega/\partial \Delta^2$ [as given in Eq.~(\ref{eq:d2F})].
 This defines the boundary between II and III.  This is substantially
 different from the line associated with $n_s(0) = 0$ (used in
 Ref.~\cite{PWY05}) which is described by
a nearly vertical
line from $(p, 1/k_Fa) = (1, 0.3)$ to $(0, 0.6)$.  Similarly the locus
of points in 
the two-dimensional parameter space $( p,
1/k_{F}a)$ where $T_c^{MF}$ vanishes defines the boundary between I and
II, as is consistent with its counterpart in Ref.~\cite{PWY05}.

In Region IID, we define $T_{unstable}$ (which is below the single
$T_c$) as the temperature where the system becomes unstable, via
Eq.~(\ref{eq:d2F}).  At a given $1/k_Fa$,
$T_{unstable}$ decreases with decreasing $p$, and approaches 0 as
$p\rightarrow 0$.  This is shown in the inset to Fig.~3 for $1/k_F a =
0.5$.  
In region IIC, the lower $T_c$ approaches 0 as
$p$ approaches 0.  Thus, there is an important distinction between the $
p \equiv 0$ and $p \rightarrow 0^+$ limits, especially at $T=0$.  For
$p$ small but finite, calculations readily encounter instabilities at
strictly $T=0$ and here superfluidity is very fragile to the
introduction of small imbalance.  By contrast, at finite $T$ this
fragility is not as pronounced.  We conclude that only $T \equiv 0$ is a
problematic temperature 
for weakly polarized superfluidity.

\begin{figure}
\centerline{\includegraphics[clip,width=3.2in]{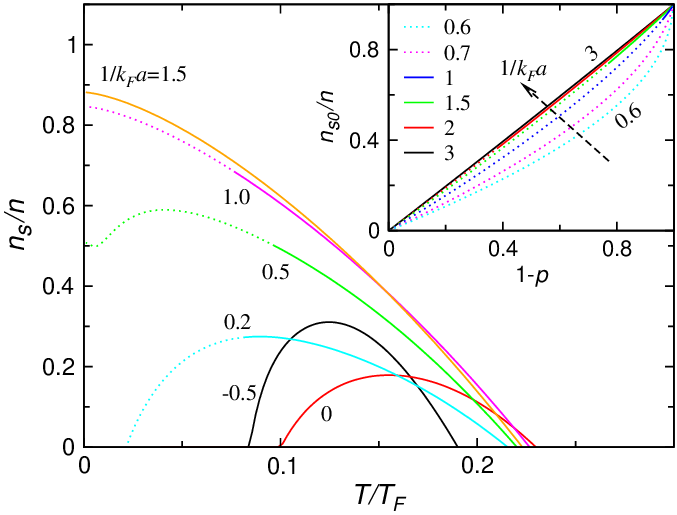}}
\caption{(Color online) Normalized superfluid density $n_s/n$ as a
  function of $T/T_F$ at $p=0.1$ for various $1/k_Fa$ from BCS to BEC,
  corresponding to regions IIC ($1/k_Fa=-0.5$ and 0), IID (0.2, 0.5,
  1.0) and III (1.5), respectively.  The inset plots $n_{s0}\equiv
  n_s(T=0)$ versus $1-p$ for different $1/k_Fa$, indicating $n_{s0}
  \rightarrow 2n_\downarrow$ in the BEC limit.  The dotted (segments of
  the) curves represent unstable solutions within region IIB (IID). }
\label{fig:ns}
\end{figure}

Figure \ref{fig:ns} presents $n_s(T)$ for $ p =0.1$.  Except for the case
$1/k_{F}a = 1.5$, Fig.~\ref{fig:ns} shows the typical behavior in
region II (of Fig.~\ref{fig:phase}), corresponding to intermediate
temperature superfluidity.  The observations here (and associated
nonmonotonicities) for $n_s(T)$ are similar in many ways to what is seen
for $\Delta(T)$ in the inset to Fig.~\ref{fig:TcMF}.  In region IIC,
$n_s$ goes to zero at the upper and lower $T_c$, whereas in IID,
$n_s(T)$ abruptly stops at $T_{unstable}$.  The dotted lines indicate
that they are in
the unstable regime.  Throughout region III, $n_s $ is found to be
monotonically decreasing with increasing $T$, as in conventional
superfluids.  Finally in the inset of Fig.~\ref{eq:n_s} we plot $n_{s0}
\equiv n_s(0)$ as a function of $1-p = 2n_\downarrow /n$.  Only in the
deep BEC regime is the dependence linear.  This plot reflects that the
excess unpaired fermions interact with the paired states, leading to a
reduced superfluid density at $T=0$ relative to $ 2 n_{\downarrow}$.

The experimental situation regarding the stability of a unitary
polarized superfluid (UPS) is currently being unraveled
\cite{PLKLH06,ZSSK06}.  If one includes the trap, within the local
density approximation it appears \cite{YD05,SM06,HS06} that the local
polarization $p(r)$, in effect, increases continuously from a small
value at the trap center to 100\% at the trap edge.  It follows from
this paper that at very low $T$ the superfluid trap center will not
support polarization, but for a range of $T$ closer to $T_c$,
polarization can penetrate the core.  We estimate from
Fig.~\ref{fig:Tc1} (assuming the central $p \approx 0.05$), that there
exists a UPS for $T \sim 0.05-0.25 T_F$.  Given the temperature range in
experiment \cite{ZSSK06} this appears to be not inconsistent with
current data ($T_F = 1.9\mu$K, $T=300\sim 505$nK $= 0.16\sim 0.27 T_F$
on resonance). In the near-BEC regime our predictions also appear
consistent with new data in Ref.~\cite{Zwierlein2006}.  More generally,
because the local
$p(r=0)$
 is small
 the unstable region is suppressed to very low $T$ as $ p \rightarrow
0$, this may explain why superfluidity in atomic traps can be observed
experimentally.  Future theory including the trap will be required to
provide quantitative comparison with experiment.

This work was supported by NSF-MRSEC Grant No.~DMR-0213745, and we thank
C. Chin, W. Yi, M.W. Zwierlein and R.G. Hulet for useful communications.


\bibliographystyle{apsrev}


\end{document}